\documentclass[prl,showpacs,twocolumn,amsmath,amssymb]{revtex4}
\usepackage{float}
\usepackage{dcolumn}
\usepackage{amsmath}
\usepackage{graphicx}
\usepackage{latexsym}
\usepackage{amsfonts}
\usepackage{amssymb}
\usepackage{bm}
\usepackage{color}

\DeclareGraphicsExtensions{.pdf,.gif,.jpg}

\def\a{\alpha}

\def\e{\epsilon}

\def\bra{\langle}
\def\ket{\rangle}

\newcommand{\be}{\begin{equation}}
\newcommand{\ee}{\end{equation}}
\newcommand{\beq}{\begin{eqnarray}}
\newcommand{\eeq}{\end{eqnarray}}

\begin{document}

\title{Time-resolved charge fractionalization in inhomogeneous 
Luttinger liquids}

\author{E. Perfetto}
\affiliation{Dipartimento di Fisica, Universit\`a di Roma
Tor Vergata, Via della Ricerca Scientifica 1, I-00133 Rome, Italy}
\affiliation{INFN, Laboratori Nazionali di Frascati, Via E. Fermi 40,
00044 Frascati, Italy}
\affiliation{European Theoretical Spectroscopy Facility (ETSF)}

\author{G. Stefanucci}
\affiliation{Dipartimento di Fisica, Universit\`a di Roma
Tor Vergata, Via della Ricerca Scientifica 1, I-00133 Rome, Italy}
\affiliation{INFN, Laboratori Nazionali di Frascati, Via E. Fermi 40,
00044 Frascati, Italy}
\affiliation{European Theoretical Spectroscopy Facility (ETSF)}

\author{ H. Kamata}
\affiliation{Department of Physics, Tokyo Institute of Technology,
2-12-1 Ookayama, Meguro, Tokyo 152-8551, Japan}

\author{T. Fujisawa}
\affiliation{Department of Physics, Tokyo Institute of Technology,
2-12-1 Ookayama, Meguro, Tokyo 152-8551, Japan}

\begin{abstract}
The recent observation of charge fractionalization
in single Tomanga-Luttinger liquids (TLLs) [Kamata et al.,
Nature Nanotech., {\bf 9} 177 (2014)] 
opens new routes for a systematic investigation of
this exotic quantum phenomenon. In this Letter we perform measurements 
on {\em two} adjacent TLLs and put forward an accurate theoretical 
framework to address the experiments. The theory is based 
on the plasmon scattering approach and can deal with injected charge pulses of arbitrary shape 
in TLL regions. We accurately reproduce and interpret the time-resolved multiple
fractionalization events  in both single and 
double TLLs. The effect of inter-correlations between the two TLLs is also 
discussed.
\end{abstract}

\maketitle

{\it Introduction.---}
When electrons are confined  in one spatial dimension
the traditional concept of Fermi-liquid quasiparticles
breaks down~\cite{giamarchi,vignale,gonzalez}.
The Fermi surface collapses and the elementary excitations become collective modes of bosonic 
nature~\cite{haldane}; these are two distinctive features of
the so-called Tomonaga-Luttinger liquid 
(TLL)~\cite{tomonaga,luttinger}.
A paradigmatic example of TLL is 
the edge state of a quantum Hall system, typically 
created on contiguous boundaries of 2D semiconductor 
heterostructures~\cite{chang}. Here the properties of the TLL
can be tuned by varying the gate voltage~\cite{kamataprb}, 
the magnetic field, the filling factor
$\nu$~\cite{chang}, and electrostatic environment of the 
channel~\cite{kumata,hashisaka}. Spatially separated 
TLLs with opposite chirality can be realized in systems with $\nu >1$ 
and, as a result of strong correlations, charge fractionalization  
occurs~\cite{lederer,lederer2}.
According to the plasmon scattering theory~\cite{safi1,safi2} an 
electron injected into a TLL region undergoes 
multiple reflections from one edge of the sample to the other. 
A fraction $r$ (dependent on the TLL parameter $g$) of the injected 
charge  $Q$
is reflected back in the adjacent edge, and the remaining fraction 
$1-r$ is transmitted forward through the same edge.
This fractionalization is a {\em transient} 
effect~\cite{safi1,safi2,salvay,psc,iucci,pcb,dolcini,oreg,neder}.
Due to charge compensations occurring at every  
fractionalization a full charge $Q$ is 
transmitted in the long-time limit.
Therefore, only time-resolved
(or finite frequency) experiments could detect the value of the
fractional charge $rQ$. 
The first conclusive evidence of transient fractionalization
was reported only recently by means of time-resolved transport
measurements of charge wave packets~\cite{kamata}. This provides a complementary 
evidence of fractionalization seen in shot-noise 
measurements~\cite{dolcini,oreg,neder,unpubl},
frequency-domain experiments~\cite{bocquillon}, and momentum-resolved 
spectroscopy~\cite{steinberg}.


In this Letter we implement the technique developed in Ref.~\cite{kamata} 
to perform transport measurements across {\em two} spatially 
separated TLLs and highlight the effect of inter-TLL interactions.
Furthemore we put forward a theoretical framework 
to calculate the evolution of wavepackets of
{\em arbitrary} shape scattering against multiple noninteracting-liquid/TLL 
interfaces arranged in different geometries. By a proper treatment of 
the boundary conditions 
we are able to make direct comparisons with the measured 
signal.
All features of the transient current are correctly captured both 
in the single and double TLL systems. 

\begin{figure}[tbp]
\includegraphics[width=0.5\textwidth]{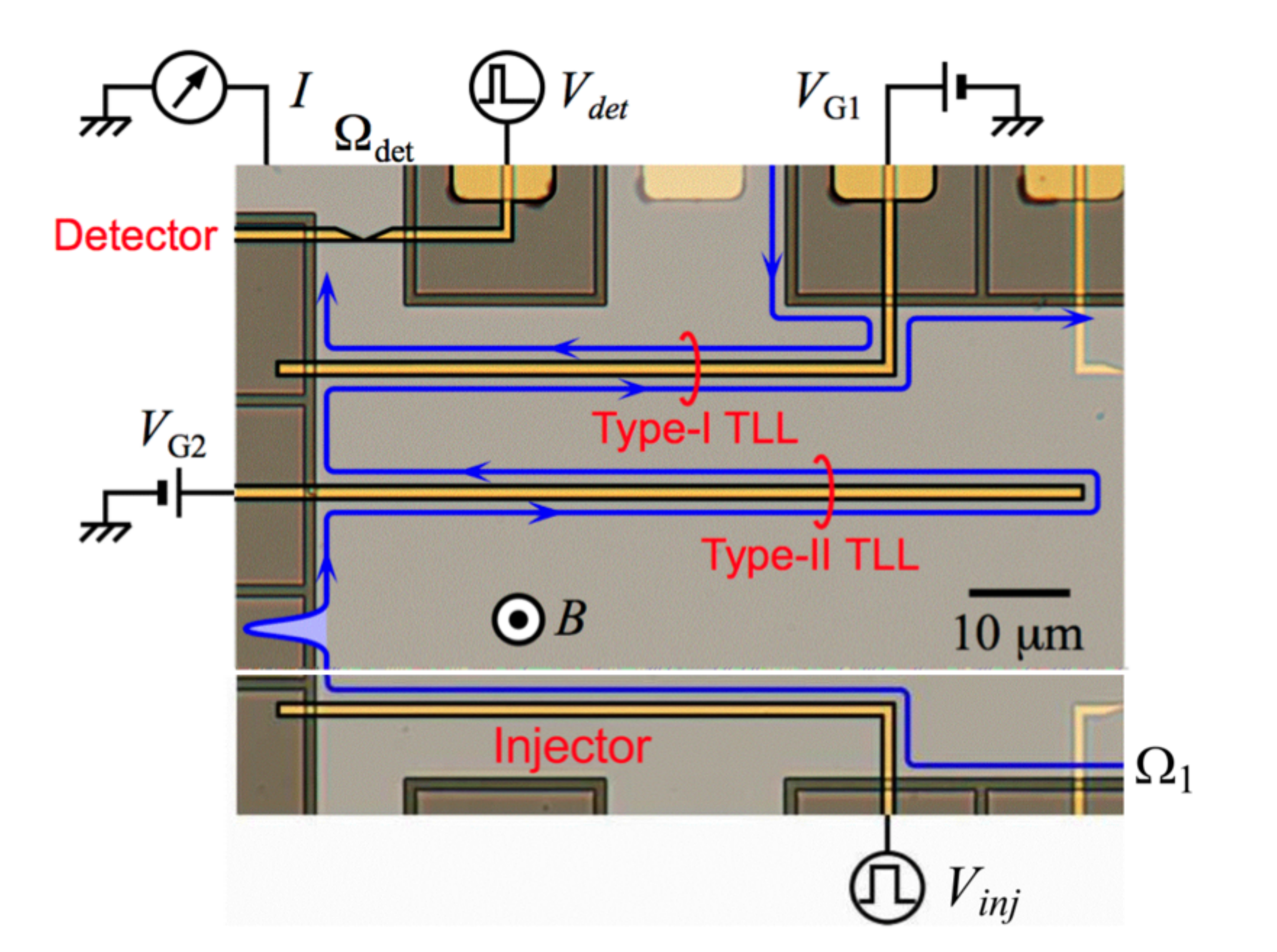}
\caption{Optical micrograph of the sample (the horizontal white line 
indicates that unused parts are not shown). Metal gate 
electrodes (gold regions) are patterned on a 2DES 
(light grey region) and etched insulating GaAs 
(dark grey regions). The 2DES located $90 \, \mathrm{nm}$ below the surface
has a density of $1.45\times 10^{11} \, \mathrm{cm}^{-2}$ and a low-temperature mobility
of $4.0\times 10^{5}\, \mathrm{ cm}^{2}  \mathrm{V}^{-1}\mathrm{s}^{-1}$. Chiral one-dimensional edge channels 
are formed along the edge of the
2DES in a strong perpendicular magnetic field $B = 4.0 \, \mathrm{T}$, which
corresponds to a bulk filling factor $\nu = 1.5$. Type-I
and Type-II TLL regions has an effective length of $\ell_{1} = 68$ and
$\ell_{2} = 80 \, \mu \mathrm{m}$,
respectively, and a width of $1 \, \mu \mathrm{m}$. The charge wavepacket is
injected at the falling edge of a voltage step $5 \, \mathrm{mV}$ in amplitude
applied to an injection gate  $V_{\mathrm{inj}}$.
The QPC detector
is set at the pinched-off regime, and one of the gate voltages is
modulated by a voltage pulse $V_{\mathrm{det}}$ of height $0.2 \mathrm{V}$ for a period of
$80 \, \mathrm{ps}$ to temporally enhance the transmission probability of the QPC. 
The average current $I$ through the QPC as a function of time interval
$t$ between two voltage pulses is measured at the detection Ohmic contact
$\Omega_{\mathrm{det}}$ under the pulse pattern repeated at $25 \, \mathrm{MHz}$. All measurements
were carried out at $\sim 300\, \mathrm{mK}$.}
\label{expdevice}
\end{figure}

{\it Experimental setup.---}
Figure \ref{expdevice} shows the sample patterned on a GaAs/AlGaAs heterostructure
with chiral one-dimensional edge channels formed along
the edge of the two-dimensional electronic system (2DES)
in a strong perpendicular magnetic field $B$. Artificial 
TLL can be formed in a pair of counter-propagating edge
channels along both sides of a narrow gate metal~\cite{kamata}. 
Other unpaired channels are considered as noninteracting
(NI) leads. Two types of TLL regions were investigated:
Type-I TLL, with NI leads on both ends,
and Type-II TLL, with NI leads only on the left and a closed
end on the right. We can selectively activate one or both the TLL
regions by applying appropriate voltages ($V_{G1}$ and $V_{G2}$). 
A non-equilibrium charge wavepacket of charge $Q\simeq 150 \;e$ is generated by depleting
electrons around an injection gate 
with a voltage
step applied on the gate. The wavepacket travels along a NI lead
as shown in Fig. \ref{expdevice}, and undergoes charge fractionalization
processes at the left and right ends of the TLL regions. The multiple
charge fractionalization processes must be investigated separately.
The reflected wavepacket appears on another NI lead, on which 
time-resolved charge detection scheme is applied with a quantum 
point contact (QPC) detector~\cite{kamataprb}. We have successfully resolved
the reflected wavepackets of charge $Q^{(\mathrm{refl})}_{1}$ 
fractionalized at the left boundary
and $Q^{(\mathrm{refl})}_{2}$ at the right boundary. Typical waveforms are shown by dots in Figs.
\ref{fig2} and \ref{fig4}. The fractionalization ratio $r$, which is related to the 
TLL parameter $g$ through $g = (1 - r)/
(1 + r)$, can be extracted from  $r = Q^{(\mathrm{refl})}_{1} / Q$ 
and is found to be approximately $g=0.92$\cite{kamata}.
The charge velocity in the TLL region can be measured from the
time interval between the two reflected wavepackets. 
The interest in activating both Type-I 
and -II regions is to assess the role of the long-range Coulomb interaction 
between the two TLLs.


{\it Model and formalism.---}
\begin{figure}[tbp]
\includegraphics[width=8.5cm]{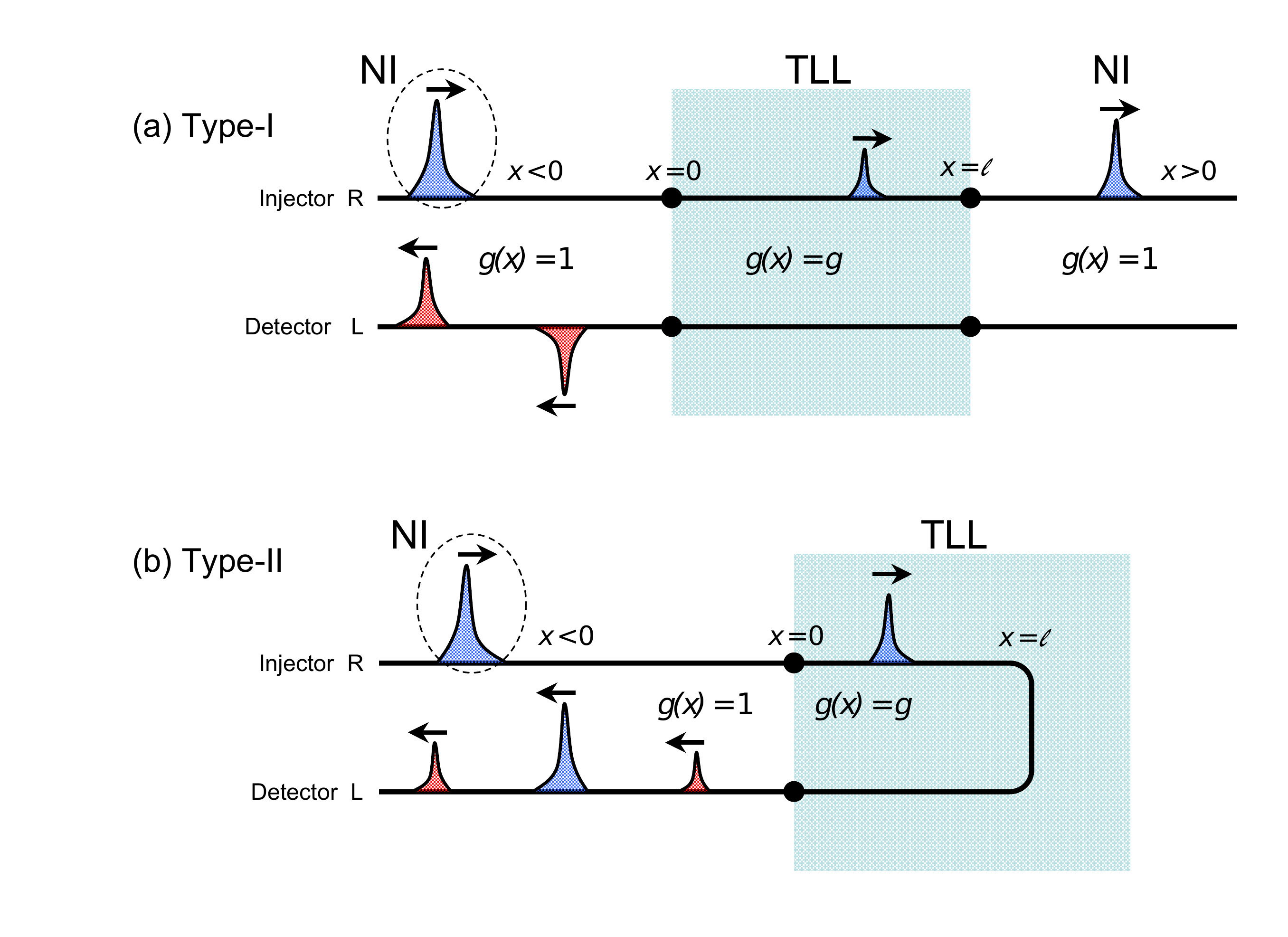}
\caption{(Color online) Model of the experimental setup. The wavepacket is injected from 
the $R$ edge (dashed circle). The figure shows a snapshot of the 
fractionalized charge when the injected wavepacket has passed the TLL 
region. Transmitted packets are dark (blue) and reflected packets are 
light (red).
Type-I geometry (a): $R$ and $L$ edges with NI regions for 
$x<0$ and $x>\ell$, and activated TLL region for 
$0<x<\ell$. 
Type-II geometry (b): a single bent edge with NI regions for $x<0$, and  
activated TLL region for $0<x<\ell$. 
}
\label{schematic}
\end{figure}
To model the setup of Fig. \ref{expdevice} we consider two parallel
chiral edges hosting  Right ($R$)  and Left ($L$) moving 
electrons, see Fig. \ref{schematic}. Electrons with opposite 
chirality experience a space-dependent repulsion $V(x)$. In the regions where
$V(x)=0$ we have a NI liquid and otherwise, $V(x)=V$,  a TLL is formed.
For electrons with the same chirality an additional 
repulsion $U(x)=U$ in the NI liquid and $U(x)=U^{\ast}$ 
in the TLL is included.
Spatial inhomogeneities in $V(x)$ induce back-scattering 
from the $R$ to the $L$ edge (and vice versa)
even without a inter-edge hopping~\cite{safi1,safi2}.
The low-energy Hamiltonian of the system reads~\cite{changRMP}
\beq
\hat{H}\!&=&\!\sum_{\a=L,R} i \a  v_{F}\int dx \,
\hat{\psi}^{\dagger}_{\a}(x)\partial_{x} \hat{\psi}_{\a}(x)   
\nonumber
\\ 
&+& \!2\pi\!\!
\int \!dx \left\{ \! V (x) \, \hat{n}_{R}(x)
\hat{n}_{L}(x)+\frac{U(x)}{2}\left[\hat{n}^{2}_{R}(x)+\hat{n}^{2}_{L}(x)\right] 
\!\right\} 
\quad \;
\label{ham}
\eeq
where the fermion field $\hat{\psi}^{(\dagger)}_{R/L}$ destroys (creates)
$R/L$ edge-state electrons moving with bare  Fermi velocity
$\a v_{F} \equiv \pm v_{F}$, and
$\hat{n}_{\a}\equiv\,\,:\hat{\psi}^{\dagger}_{\a} \hat{\psi}_{\a}:$ is 
the density fluctuation operator. For a nonperturbative treatment of
the interaction  we
bosonize the field operators as
$
\hat{\psi}_{\a}(x)=\frac{\eta_{\a}}{\sqrt{2\pi a}}
e^{-2\sqrt{\pi}\,i\hat{\phi}_{\a}(x)},
\label{bospsi}
$
with
$\eta_{\a}$ the anticommuting Klein factor, $a$ a short-distance 
cutoff, and  $\hat{\phi}_{\a}(x)$ the chiral boson fields. The   
density can then be expressed as
$
\hat{n}_{\a}=-\partial_{x}\hat{\phi}_{\a}/\sqrt{\pi}.
$
By introducing the auxiliary fields  
$\hat{\phi}=\hat{\phi}_{L}+\hat{\phi}_{R}$
and  $\hat{\theta}=\hat{\phi}_{L}-\hat{\phi}_{R}$
 Eq. (\ref{ham}) becomes~\cite{giamarchi}
\be
\hat{H}=\frac{1}{2}\int dx  \left\{ \frac{v(x)}{g(x)} [\partial_{x} 
\hat{\phi}(x)]^{2} + v(x)g(x)  [\partial_{x} 
\hat{\theta}(x)]^{2} \right\}, 
\ee
where for a TLL region of length $\ell$ the
parameter $g(x)$ and the renormalized velocity 
$v(x)$ depends on the interactions through the relations
\beq
g(x)&=&
\left\{
\begin{array}{ccc}
  \sqrt{\frac{v_{F}+U^{*}-V}{v_{F}+U^{*}+V}} \equiv g &\quad&  \mathrm{for}
  \quad 0<x<\ell   \\
1 &\quad&  \mathrm{otherwise} \\
\end{array}
 \right. \nonumber \\
 \label{gandv} \\
 v(x)&=&
\left\{
\begin{array}{ccc}
  \sqrt{(v_{F}+U^{*})^{2}-V^{2}} \equiv v^{*} &\quad&  \mathrm{for}  
  \quad 0<x<\ell   \\
v_{F}+U \equiv v &\quad&  \mathrm{otherwise.}  \nonumber \\
\end{array}
 \right.
\eeq

The temporal evolution of the system is governed by the equation
of motion for $\hat{\phi}$~\cite{suppmat}. Taking the average $\phi(x,t)\equiv \bra 
\hat{\phi}(x,t) \ket$ over an arbitrary wavepacket state we find 
%
\be
\frac{d^{2}}{dt^{2}}\phi(x,t)=v(x)g(x)
\partial_{x}\left(\frac{v(x)}{g(x)} 
\partial_{x}\phi(x,t)\right),
\label{eqdiff}
\ee
which implies that $\phi$ and  $\frac{v(x)}{g(x)} 
\partial_{x}\phi$ are continuous for all $x$. 
For independent channels, as those of the Type-I geometry illustrated in 
Fig.~\ref{schematic}, these
are the only conditions to impose on the solution of Eq. 
(\ref{eqdiff})~\cite{safi1,hashisaka,HRHL.2011}. On the other hand, for the 
Type-II geometry  one has to further impose that $R$ electrons are 
converted into $L$ electrons and viceversa, i.e., that the channels
are not independent. The proper treatment of boundary conditions, 
absent in previous works,
leads to a qualitative different 
transient fractionalization since the transmission and reflection 
coefficients are entangled.   
Once $\phi(x,t)$ is known the total 
density and current  are extracted from $ \rho(x,t) = e \bra \hat{n}(x,t) \ket = 
-e\partial_{x}\phi(x,t)/\sqrt{\pi}$ and $j(x,t)= e  
\partial_{t} \phi (x,t)/\sqrt{\pi}$.

We consider an incident  wavepacket injected in 
the upper $R$ edge, see Fig.~\ref{schematic}. Then the solution of 
Eq.~(\ref{eqdiff}) can be expanded in right-going scattering states $s_{q}(x)$ of 
energy $\e_{q}=vq$ according to $\phi(x,t)=\int_{-\infty}^{\infty} \frac{dq}{2\pi} 
\phi_{q}s_{q}(x)e^{-i\e_{q}t}$~\cite{bcnote}. For a wavepacket 
initially, say at time $t=0$, localized in $x<0$ the function
$\phi_{q}$ is related to the Fourier transform $\rho^{(\mathrm{inc})}_{q}$ of 
$\rho^{(\mathrm{inc})}(x)=\rho(x,0)$ 
by the relation $\phi_{q}=\frac{i\sqrt{\pi}}{eq}
\rho^{(\mathrm{inc})}_{q}$~\cite{nontrivial}.
Therefore, once $s_{q}(x)$ is known the time-dependent density and
current are given by
\beq
\rho (x,t)&=&-i \int_{-\infty}^{\infty} \frac{dq}{2\pi}
\frac{\rho^{(\mathrm{inc})}_{q}}{q}  e^{-i\e_{q}t} 
\partial_{x}s_{q}(x), \nonumber \\
j (x,t)&=& v \int_{-\infty}^{\infty} \frac{dq}{2\pi}
\rho^{(\mathrm{inc})}_{q}  e^{-i\e_{q}t} s_{q}(x).
\label{curreq}
\eeq
Below we solve the scattering problem in 
the geometries of the experiment.

{\it Type-I geometry.---} 
This geometry is illustrated in Fig. \ref{schematic}.a and has been 
realized in Ref.~\cite{kamata}. We look for scattering 
states of the form
\be
s_{q}(x)=
\left\{
\begin{array}{ccc}
  e^{iqx} + r_{q} e^{-iqx}      & \quad  \mathrm{for} &  \quad x<0   \\
a_{q}e^{iq'x} + b_{q} e^{-iq'x} & \quad  \mathrm{for} &  \quad 0<x<\ell   \\
t_{q}e^{iqx}                   &\quad  \mathrm{for}  &  \quad x>\ell 
\, ,   
\end{array}
 \right.
\ee
with $q'=\frac{v}{v^{*}}q$. By imposing the continuity conditions at 
the boundaries
we obtain a $4 \times 4$ linear system~\cite{suppmat}
that we solve exactly.
If we are interested in the current detected at 
the collector (located in $x<0$) only the reflection coefficient $r_{q}$ is 
needed~\cite{nota2}: 
\be
r_{q}=-r+4g\sum_{n=1}^{\infty}\zeta_{n}e^{2inq'\ell},
\label{refl}
\ee
where $g_{\pm} =1\pm g$, $r=\frac{g_{-}}{g_{+}}$, and
$\zeta_{n}=\frac{g_{-}^{2n-1}}{g_{+}^{2n+1}}$. Inserting this 
expression in Eq. (\ref{curreq}) 
the time-dependent density and current for $x<0$ read
\beq
\rho(x,t)&=&\rho^{(\mathrm{inc})}(x_{-})+\rho^{(\mathrm{refl})}(x_{+}) 
\nonumber \\
j(x,t)&=&v[\rho^{(\mathrm{inc})}(x_{-})-\rho^{(\mathrm{refl})}(x_{+})], 
\label{solution}
\eeq
with $x_{\pm}=x\pm vt$, $x_{n}=\frac{2n\ell v}{v^{*}}$, and
\be
\rho^{(\mathrm{refl})}(x_{+})=r
\rho^{(\mathrm{inc})}(-x_{+}) 
- 4g\sum_{n=1}^{\infty}\zeta_{n} \rho^{(\mathrm{inc})}(-x_{+}+x_{n}).
\label{solution1}
\ee
Equation (\ref{solution1}) generalizes the result of Ref.~\cite{safi1} to arbitrary wavepacket shapes.
The first reflection occurs at 
time $t_{1}= |x_{0}|/v$ ($x_{0}<0$ being the initial position of the 
wavepacket) at the left boundary and a fractionalized charge 
$Q^{(\mathrm{refl})}_{1}=rQ$ is reflected back in the 
 $L$ edge (here $Q=\int dx \rho^{(\mathrm{inc})}(x)$). 
The transmitted fractional charge propagates in the TLL region, 
a second reflection occurs at the right boundary and  
at time $t_{2}=t_{1}+2\ell/v^{*}$ a second wavepacket of charge
$Q^{(\mathrm{refl})}_{2}=-Q(4 gg_{-}/
g_{+}^{3})=-Qr(1-r^{2})$ appears in the $L$ edge. 
The fractionalization sequence continues {\em ad infinitum} 
and  the reflected charge $Q^{(\mathrm{refl})}_{n}$ 
diminishes at each event.
At the end of the infinite 
sequence the total reflected charge vanishes 
since $Q^{(\mathrm{refl})}=\sum_{n=1}^{\infty}Q^{(\mathrm{refl})}_{n}=
-r-4g\sum_{n=1}^{\infty}\zeta_{n}=0$. This is a consequence of the 
chiral charge conservation and highlights the transient nature of the 
fractionalization phenomenon.
\begin{figure}[tbp]
\includegraphics[width=0.4\textwidth]{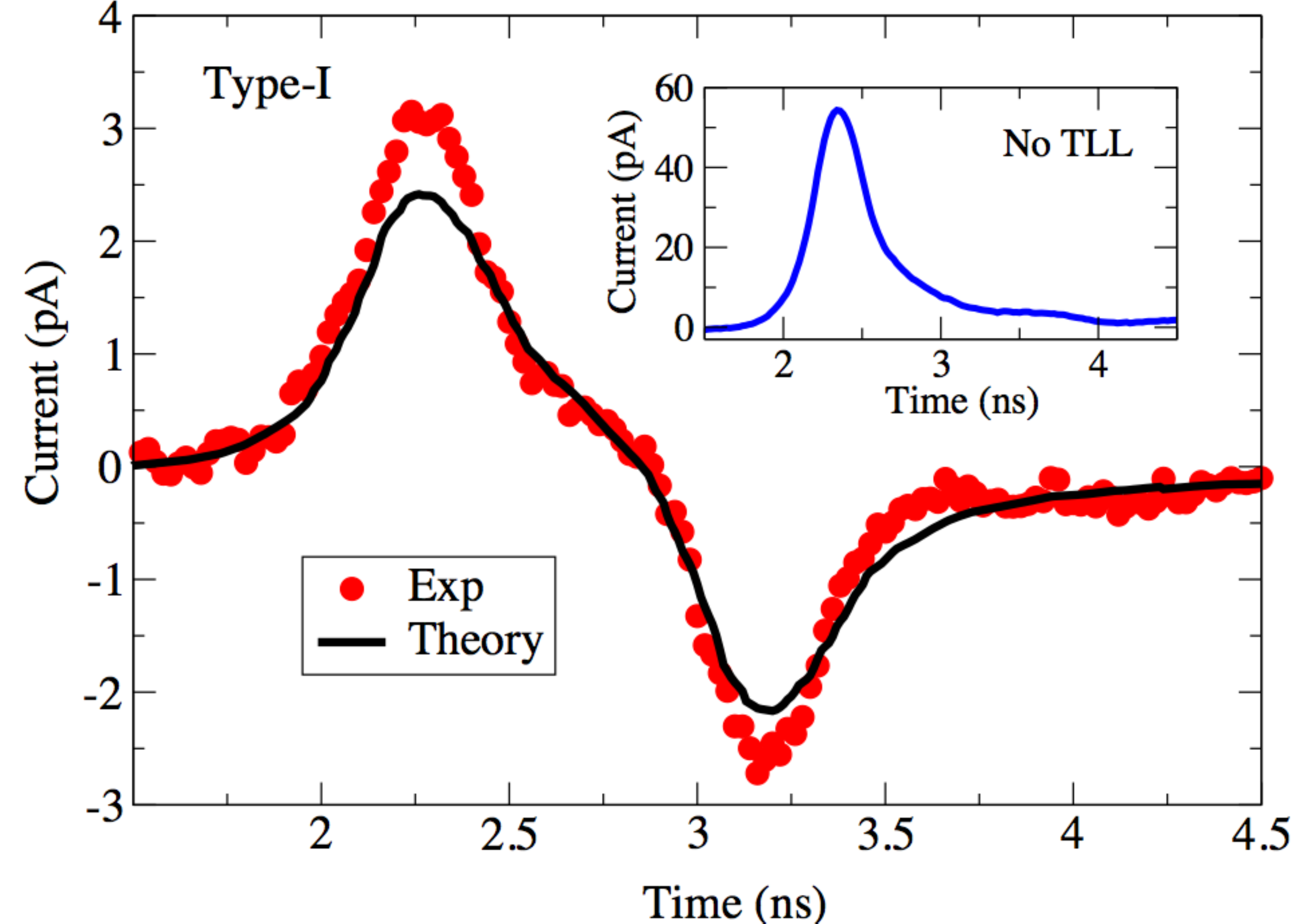}
\caption{Type-I geometry: Calculated current (black curve) 
from Eqs. (\ref{solution},\ref{solution1}) 
versus  measured current (dotted-red curve) from Ref. 
\cite{kamata}. The inset 
shows the incident waveform $v\rho^{(\mathrm{inc})}$.}
\label{fig2}
\end{figure}
For the comparison with the experiment we acquire 
$\rho^{(\mathrm{inc})}(x_{0}-vt)$ from Ref. \onlinecite{kamata},
see inset in  Fig. \ref{fig2} and used  
$g=0.92$, $\ell=\ell_{1}=68$ $\mu$m, $v^{*}=150$ km/s
and estimated $v$ by a best fitting.
As shown in  Fig. \ref{fig2} the agreement with the current calculated from Eq. 
(\ref{solution}) 
is remarkably good.


{\it Type-II geometry.---}
Here a single edge is bent on itself as illustrated in Fig. 
\ref{schematic}.b. Therefore $R$ electrons in the upper branch  
are converted in $L$ electrons
in the lower branch. We model this geometry by imposing that the $L$ 
amplitude $b_{q}$ of the scattering state in the TLL 
region equals $-a_{q}e^{2iq'\ell}$~\cite{suppmat}.
Following the same line of reasoning as before we find
the reflection coefficient
\beq
r_{q}=-r+4g\sum_{n=1}^{\infty}\xi_{n}e^{2inq'\ell},
\label{refl2}
\eeq
with $\xi_{n}=(-1)^{n}\frac{g_{-}^{n-1}}{g_{+}
^{n+1}}$.
We observe that $|r_{q}|=1$ as it should due to charge conservation. 
The density and current at the collector in
$x<0$ are still given by Eq. (\ref{solution}) but the reflected 
density reads
\be
\rho^{(\mathrm{refl})}(x_{+})=r
\rho^{(\mathrm{inc})}(-x_{+}) 
-4g\sum_{n=1}^{\infty}\xi_{n} \rho^{(\mathrm{inc})}(-x_{+}+x_{n}).
\label{solution2}
\ee
\begin{figure}[tbp]
\includegraphics[width=0.4\textwidth]{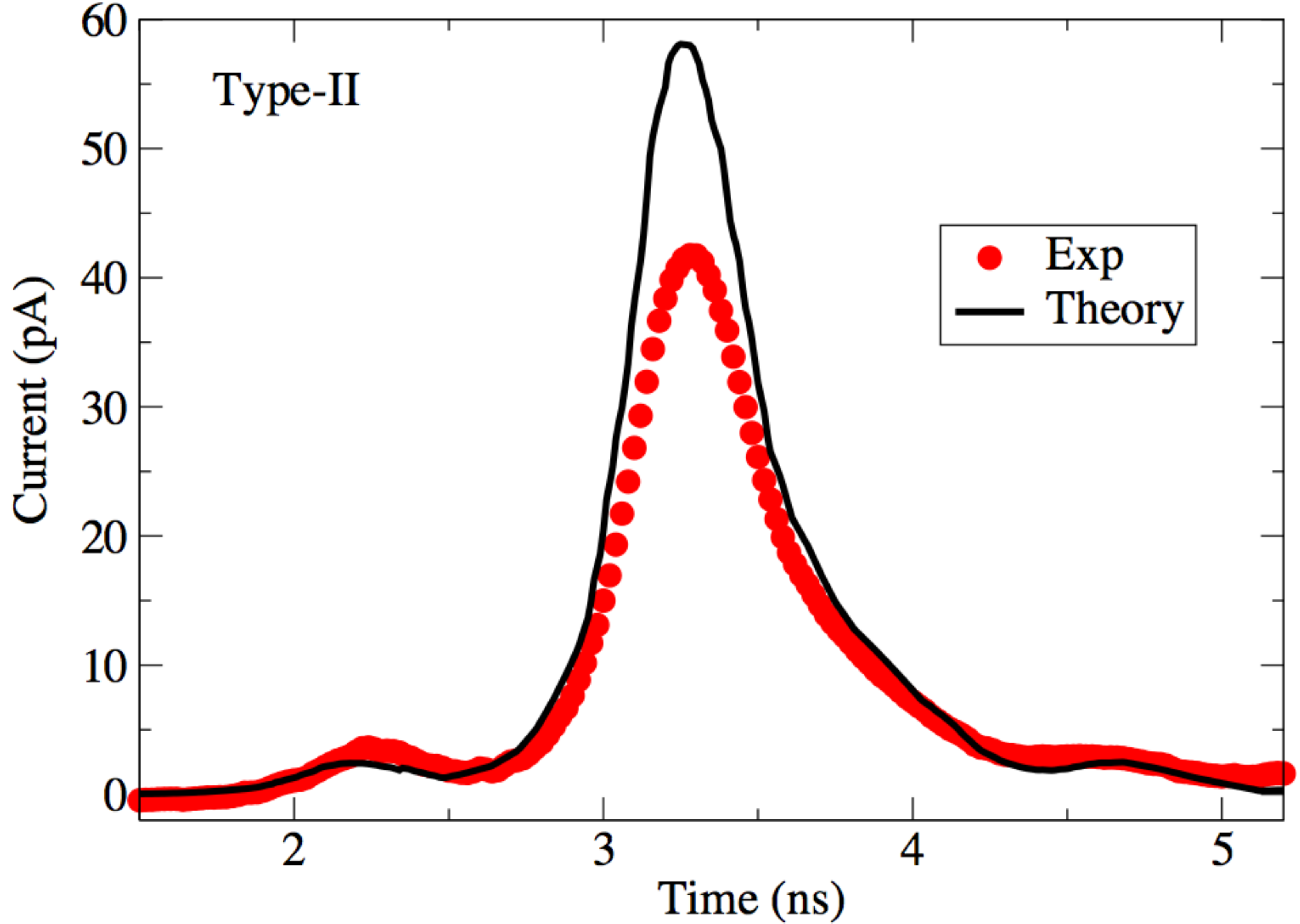}
\caption{Type-II geometry: Calculated current (black curve) 
from Eqs. (\ref{solution},\ref{solution2}) 
versus measured current (dotted-red  curve) from Ref. 
\cite{kamata}.}
\label{fig4}
\end{figure}
In Fig. \ref{fig4} we show the calculated (black curve)
and measured~\cite{kamata} (dotted-red curve)
current in the lower branch. The parameters are the same as in 
Fig. \ref{fig2} with the only 
difference that  $\ell=\ell_{2}=80$ $\mu$m.
Again a good agreement between theory and experiment 
is found. The theory reproduces a small 
first reflection of charge $rQ$ 
(occurring at time $t_{1}$) and a 
subsequent large transmitted charge $(4g/g_{+}^{2})Q$ 
(occurring at time $t_{2}$).

\begin{figure}[tbp]
\includegraphics[width=0.41\textwidth]{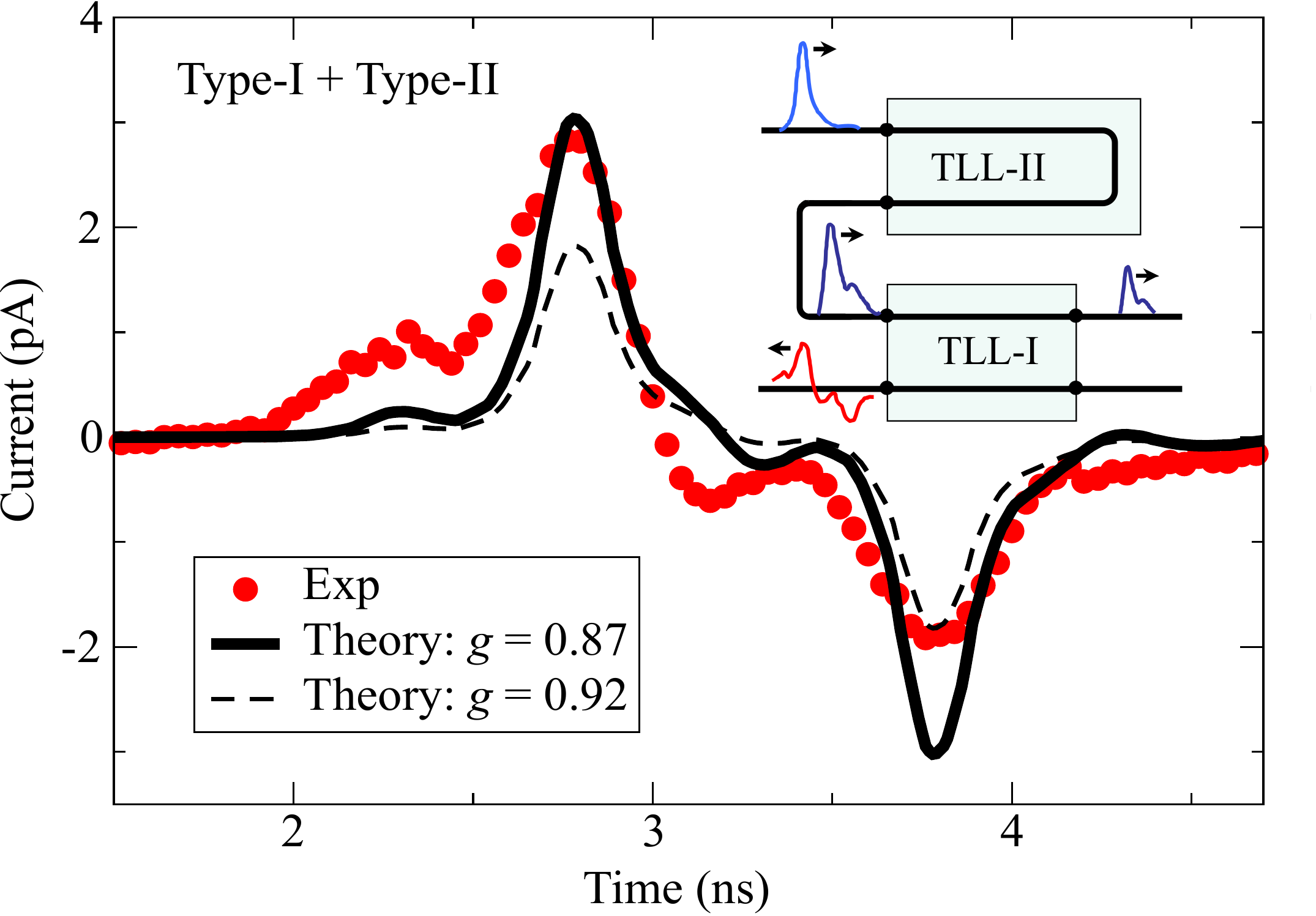}
\caption{
Measured current (dotted-red curve) from TLL-I when both TLL-I and TLL-II are 
activated versus calculated current with $g=0.92$ 
(black-dashed curve) and $g=0.87$ (black-solid curve). 
The inset shows a cartoon of the fractionalization process.
The velocities in TLL-I and TLL-II are different (with  $V_{G1} = 
-0.19\, \mathrm{V}$ and $V_{G2} = -1.4\, \mathrm{V}$) 
in order to isolate four fractionalized wavepackets.
}
\label{fig7}
\end{figure}

{\it Type-I + Type-II geometry.---}
Finally we present numerical and experimental results 
when both Type-I and Type-II TLLs 
are activated. As illustrated in Fig. \ref{expdevice} 
the wavepacket injected into TLL-II is  partially transmitted
toward TLL-I and the resulting reflected wavepacket 
is then measured at the collector.
The measured signal is displayed in Fig. \ref{fig7}  (dotted red 
curve).
The simultaneous activation of TLL-I and TLL-II produces a richer 
current pattern characterized by an  
additional peak and dip.
These extra 
structures are naturally interpreted within our theory.
The reflected wavepacket is given by 
$\rho^{(\mathrm{refl})}(x_{+})$ with only TLL-I 
activated by replacing $\rho^{(\mathrm{inc})}(x)$ in Eq. 
(\ref{solution})  
with the outcome $\rho(x_{-})$ obtained by a preliminary calculation 
with only TLL-II activated.
TLL-II alone produces a waveform similar to the 
incident one, with the addition of a small side peak of weight $r$ on 
the left, see Fig. \ref{fig4}. 
The temporal delay between the peaks is $\Delta t_{II}=2 
\ell_{2}/v^{*}_{II}$, where $v^{*}_{II}$ is the renormalized velocity 
inside TLL-II.
When this double-peaked wavepacket enters TLL-I the reflected current displays 
a first replica of the incident shape with positive weight $r$ and a 
second replica of the incident shape with negative weight $-r(1-r^{2})$, as we 
demonstrated in Fig. \ref{fig2}. 
The delay between the two replicas is $\Delta t_{I}=2 
\ell_{1}/v^{*}_{I}$, $v^{*}_{I}$ being the renormalized velocity 
inside TLL-I. This explains the 
experimentally observed pattern of Fig. \ref{fig7} (the inset 
shows a cartoon of this double fractionalization process).


The calculated reflected current is shown in Fig. \ref{fig7} 
for comparison. From $\Delta 
t_{I(II)}=\ell_{1(2)}/v^{*}_{I(II)}$
with $\Delta t_{I}\approx 1.0$ ns and $\Delta 
t_{II}\approx 0.5$ ns we estimated 
$v^{*}_{I} \approx 136$ km/s, $v^{*}_{II} \approx 320$ km/s,
and $v$ by a best fitting. 
The value  $g=0.92$ (black-dashed curve) is probably
too large as the additional peak and dip are almost 
invisible. We therefore repeated the calculation with
$g = 0.87$
(black-solid curve) to match the height of the positive main peak
and found that the additional peak and dip are correctly more 
pronounced. The physical justification of a smaller $g$ is elaborated in 
the conclusions.

{\it Conclusions.---}
We extended the plasmon scattering approach to address 
the charge fractionalization phenomenon
recently observed in artificial TLLs of different geometries~\cite{kamata}.
The method allows us to monitor the temporal evolution
of a charge wavepacket in each chiral edge of the experimental setup,
thus providing a tool for a direct comparison with the time-resolved 
transport measurement.
Quantitative agreement between theory and experiment is obtained for 
the Type-I and Type-II geometries. We then performed new measurements 
in a double-TLL geometry and found indications   
that electron correlations are enhanced due to the repulsion between 
electrons in different TLLs. 
Our calculations neglect the inter-TLL repulsion and the enhancement 
of correlations is 
effectively accounted for by a reduced TLL parameter $g$.
The proper inclusion of the
long-range interaction across the bulk 2DEG is eventually required
for the ultimate understanding
of the transport properties of interacting edge channels. 


E.P. and G.S. acknowledge funding by MIUR FIRB 
grant No. RBFR12SW0J.
H.K. and T.F. acknowledge funding by JSPS KAKENHI (21000004, 11J09248).
We also thank N. Kumada, M. Hashisaka, and K. Muraki for experimental supports.


\begin{center}
\begin{figure}[tbp]
\includegraphics[width=1.\textwidth]{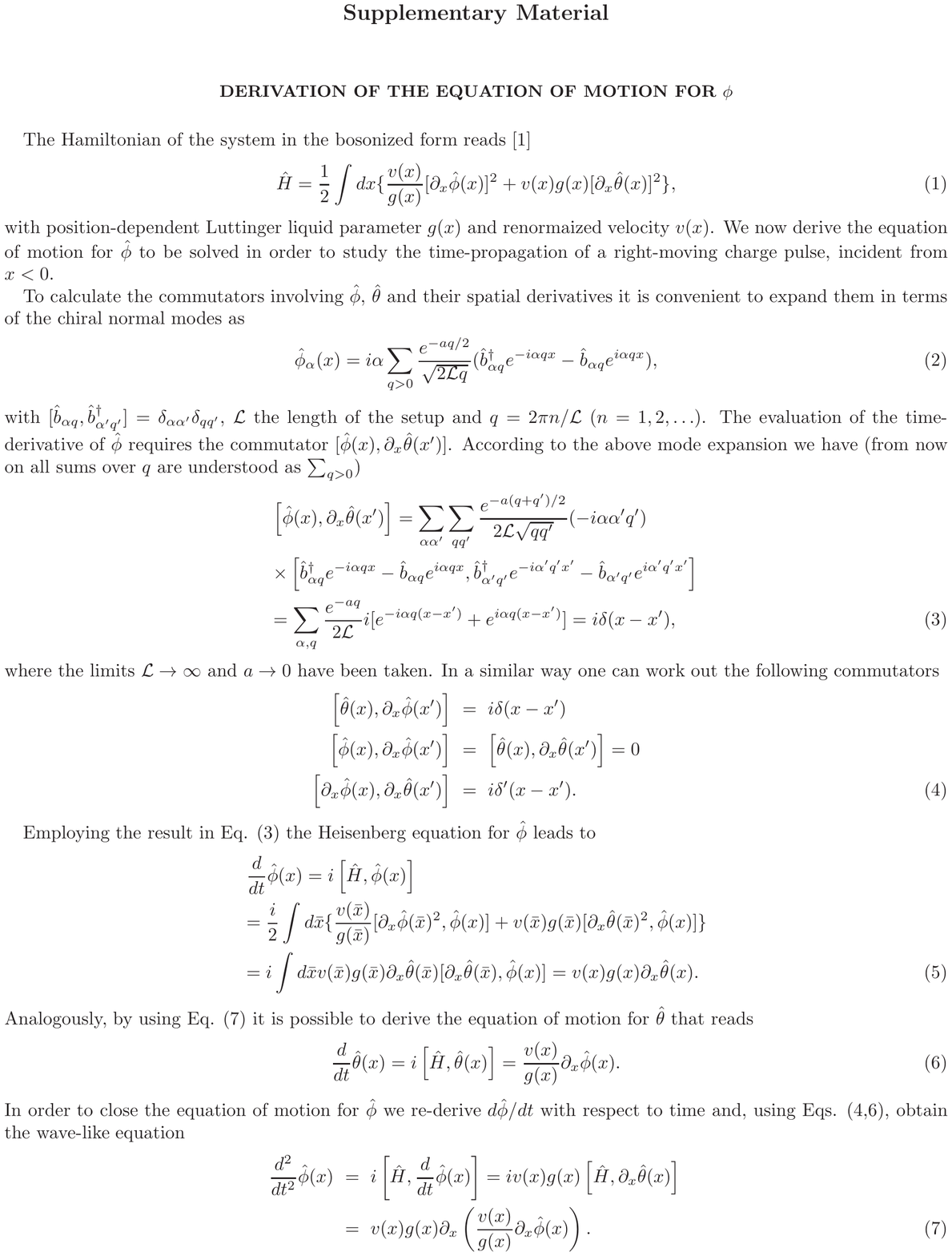}
\end{figure}
\end{center}

\begin{center}
\begin{figure}[tbp]
\includegraphics[width=1.\textwidth]{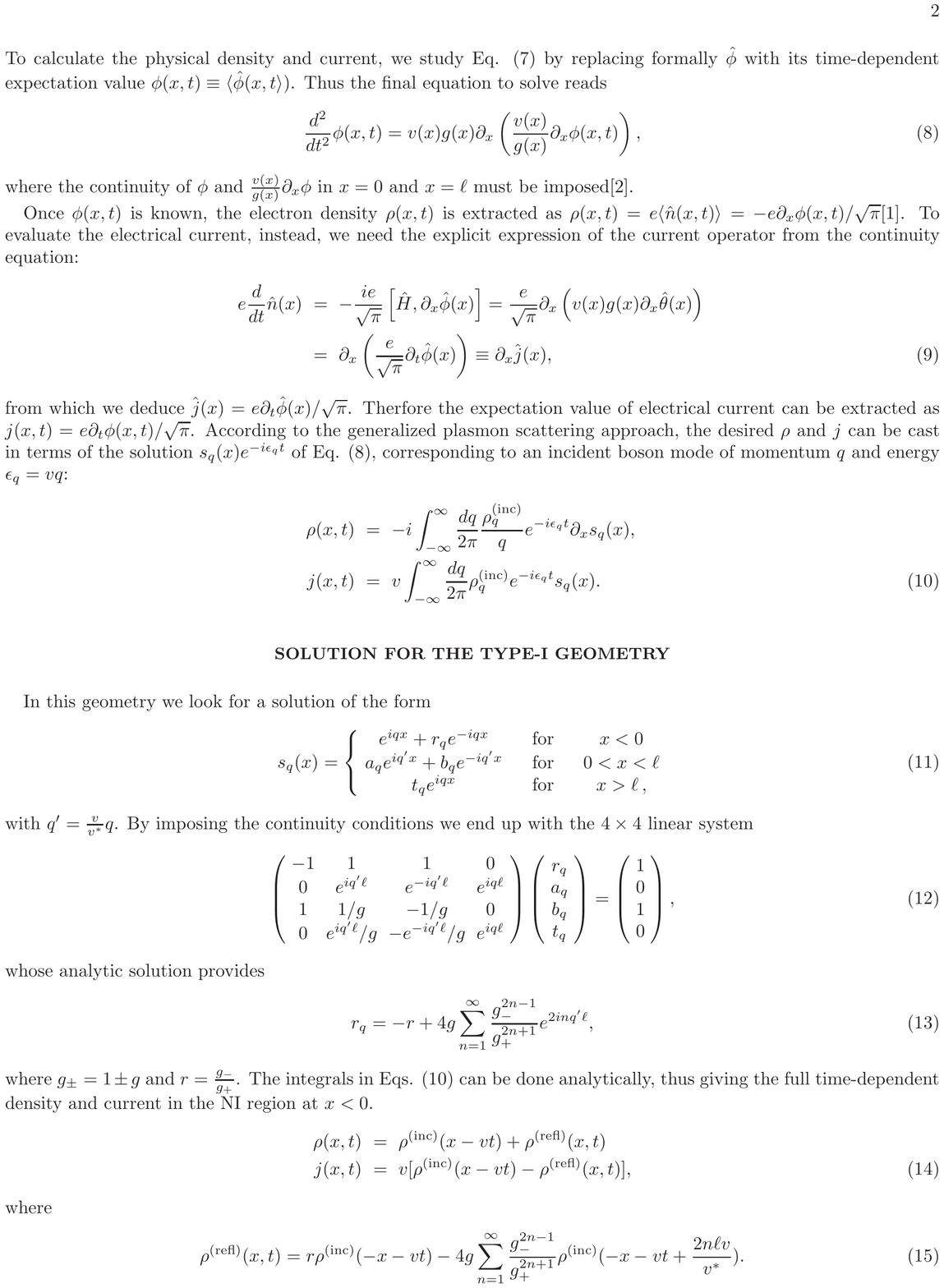}
\end{figure}
\end{center}

\begin{center}
\begin{figure}[tbp]
\includegraphics[width=1.\textwidth]{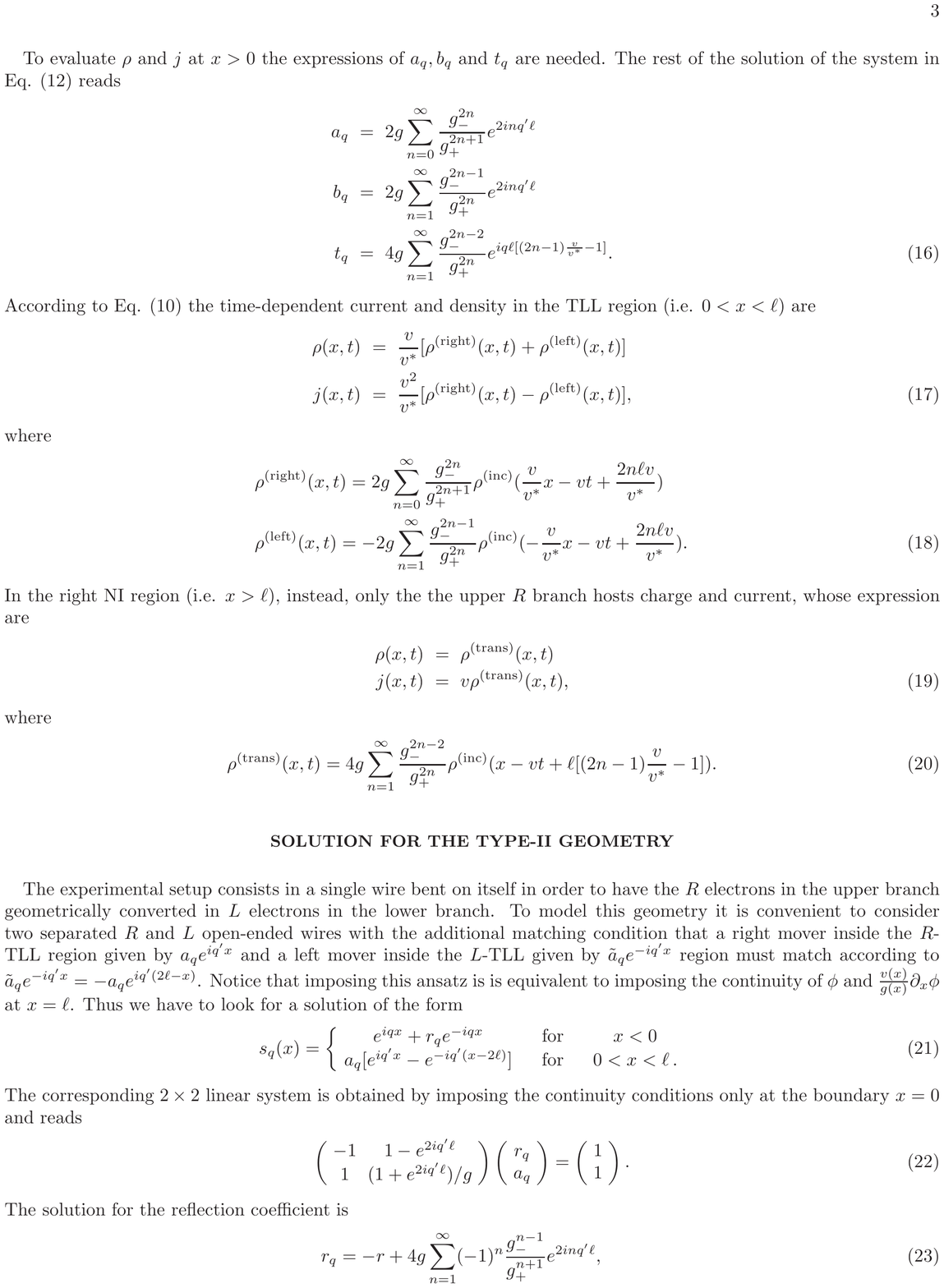}
\end{figure}
\end{center}

\begin{center}
\begin{figure}[tbp]
\includegraphics[width=1.\textwidth]{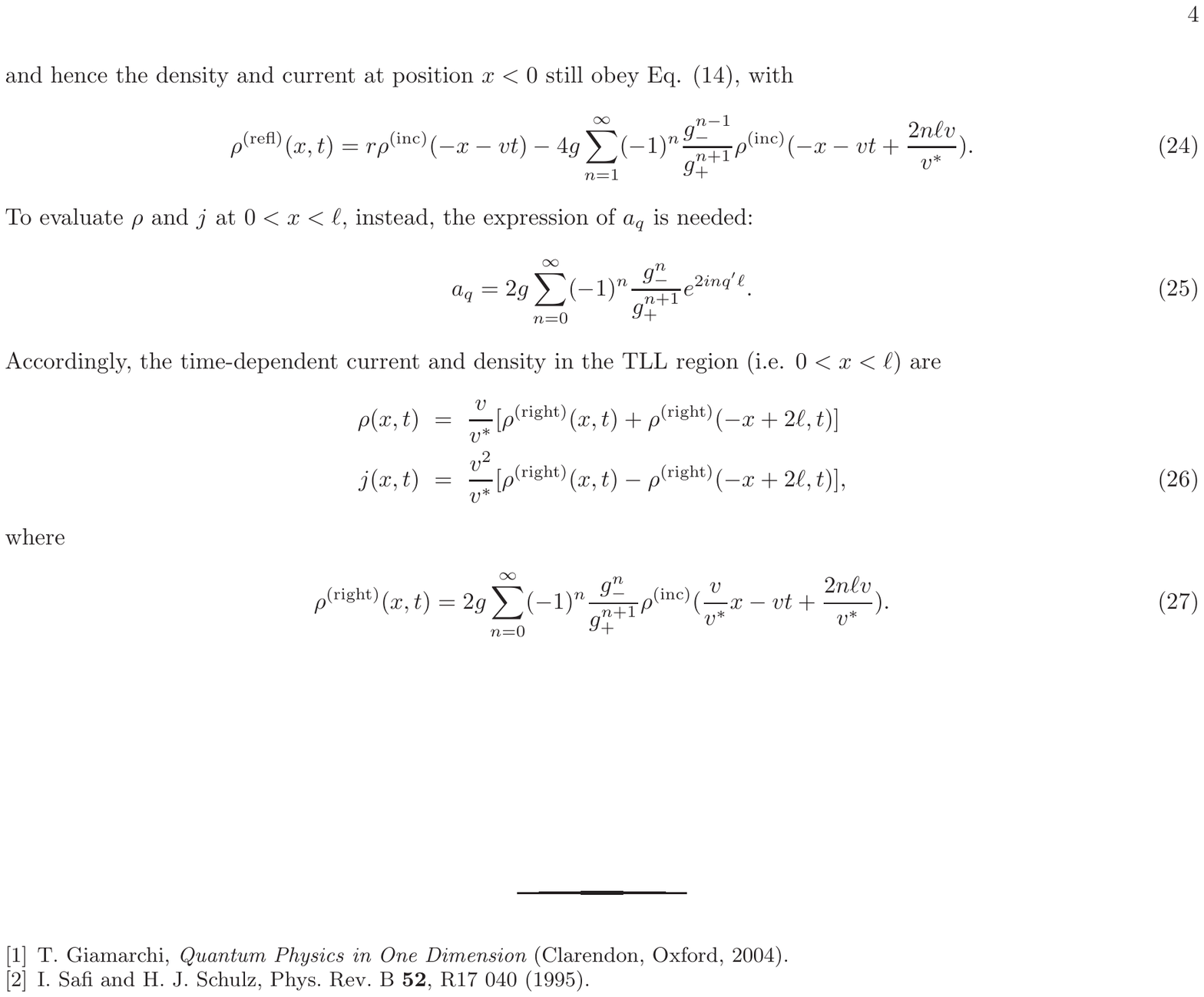}
\end{figure}
\end{center}


\begin{thebibliography}{99}
\bibitem{nontrivial}
The property that the expansion coefficients of 
$\rho^{(\mathrm{inc})}(x)$ are the same in the scattering-state basis 
and in the plane-wave basis is crucial to perform the $q$-integral in 
Eqs. (\ref{curreq}). This property can be checked by calculating 
$\rho^{(\mathrm{inc})}(x)$ for all $x$, see Ref.~\onlinecite{suppmat}.
\bibitem{nota2}
To evaluate $\rho$ and $j$ in $x>0$ the expressions of 
$a_{q},b_{q}$ and $t_{q}$ are needed~\cite{suppmat}.
\end{thebibliography}
\end{document}